\documentclass{iau} 
\usepackage{graphicx}
\usepackage{hyperref}

\title[Near-core magnetic field inference of HD~43317] 
{Asteroseismology reveals the near-core magnetic field strength in the \\ early-B star HD~43317}

\author[D. M. Bowman et al.]   
{D. M. Bowman$^{1,2}$, D. Lecoanet$^{3,4,2}$ \and T. Van Reeth$^{1,2}$}

\affiliation{
$^1$Institute of Astronomy, KU Leuven, Celestijnenlaan 200D, 3001 Leuven, Belgium \\ email: {\tt dominic.bowman@kuleuven.be} \\[\affilskip]
$^2$Kavli Institute for Theoretical Physics, University of California, \\ Santa Barbara, CA 93106, USA \\
$^3$Department of Engineering Sciences and Applied Mathematics, \\ Northwestern University, Evanston, IL 60208, USA \\
$^4$CIERA, Northwestern University, Evanston, IL 60201, USA \\

}

\pubyear{2022}
\volume{361}  
\jname{Massive Stars Near and Far}
\editors{N. St-Louis, J. S. Vink \& J. Mackey, eds.}
\begin{document}

\maketitle

\begin{abstract}
Spectropolarimetic campaigns have established that large-scale magnetic fields are present at the surfaces of approximately 10\% of massive dwarf stars. However, there is a dearth of magnetic field measurements for their deep interiors. Asteroseismology of gravity-mode pulsations combined with rotating magneto-hydrodynamical calculations of the early-B main-sequence star HD~43317 constrain its magnetic field strength to be approximately $5 \times 10^5$~G just outside its convective core. This proof-of-concept study for magneto-asteroseismology opens a new window into the observational characterisation of magnetic fields inside massive stars.

\end{abstract}

\firstsection 

\section{Introduction}

Only about 10\% of early-type (i.e. spectral types O, B and A) dwarf stars are known to host a strong, large-scale magnetic field at their surface, which is thanks to dedicated spectropolarimetric campaigns \cite[(Wade et al. 2014; Grunhut et al. 2017; Shultz et al. 2019; Sikora et al. 2019)]{Wade2014, Grunhut2017, Shultz2019, Sikora2019}. The magnetic fields are predominantly dipolar, typically inclined with respect to the rotation axis, and have strengths between approximately 100~G and a few tens of kG. Large-scale magnetic fields in main-sequence stars are thought to be formed through the star formation process \cite[(e.g. Neiner et al. 2015)]{Neiner2015} or through binary star mergers that leave behind a strongly magnetic and apparently single star \cite[(e.g. Schneider et al. 2019)]{Schneider2019}. Regardless of how they arise, magnetic fields are one of the largest uncertainties in stellar structure and evolution theory, as they strongly impact interior rotation, mixing and angular momentum transport \cite[(Maeder \& Meynet 2005; Keszthelyi et al. 2019)]{Maeder2005, Keszthelyi2019}.

However, there are few inferences of the strength and geometry of a magnetic field below the surface of an early-type star, specifically in the deep interior near the convective cores of main-sequence massive stars. Thus any direct inference of interior stellar properties from asteroseismology -- the study of stellar structure from pulsations -- are extremely valuable in improving stellar structure and evolution theory \cite[(Aerts 2021)]{Aerts2021}. To date, only a handful of massive stars have undergone forward asteroseismic modelling to constrain their interior properties (see \cite[Bowman 2020]{Bowman2020} for a review). 

Amongst massive stars, there are two main types of pulsations that can be excited: pressure and gravity modes, which are standing waves that are restored by the pressure force and buoyancy, respectively. Coherent pulsation modes (i.e. standing waves) are excited by a heat-engine mechanism operating in the iron-nickel opacity bump at 200\,000~K, and have pulsation periods of order hours to days \cite[(Dziembowski \& Pamyatnykh 1993; Dziembowski et al. 1993)]{Dziembowski1993a, Dziembowski1993b}. Gravity modes in particular are sensitive to the properties (e.g. mass and size) of the convective core in main-sequence massive stars, because they probe the Brunt-V{\"a}is{\"a}l{\"a} frequency profile. Furthermore, rotation shifts pulsation mode frequencies such that the interior rotation profiles of massive stars can be deduced from gravity-mode period spacing patterns \cite[(Bouabid et al. 2013)]{Bouabid2013} and rotationally split multiplets \cite[(e.g. Aerts et al. 2003)]{Aerts2003}.

In these proceedings, we summarise the methods and results of applying magneto-asteroseismology to the magnetic, pulsating, early-B star HD~43317 published by \cite[Lecoanet, Bowman \& Van Reeth (2022)]{Lecoanet2022}, to which the reader is referred for full details.

\section{The Magnetic Pulsator HD~43317}

HD 43317 is an early-type main-sequence star with spectral type B3.5\,V. From dedicated high-resolution spectropolarimetry, a precise rotation period of  0.897673(4)~d and a large-scale dipolar surface magnetic field with strength $B_{\rm p} = 1312 \pm 332$~G have been measured \cite[(Briquet et al. 2013; Buysschaert et al. 2017)]{Briquet2013, Buysschaert2017}. Furthermore, atmospheric modelling demonstrated that HD~43317 is likely to be a single star with solar metallicity, and has an effective temperature of $T_{\rm eff} = 17350 \pm 750$~K, a surface gravity of $\log\,g = 4.0 \pm 0.1$ and a projected surface rotational velocity of $v\,\sin\,i = 115 \pm 9$~km\,s$^{-1}$ \cite[(P{\'a}pics et al. 2012)]{Papics2012}.

HD~43317 was observed by the CoRoT space mission \cite[(Auvergne et al. 2009)]{Auvergne2009}, which assembled a light curve spanning 150.5~d with an average cadence of 32~s. Frequency analysis of the CoRoT light curve revealed dozens of significant gravity-mode frequencies \cite[(P{\'a}pics et al. 2012; Buysschaert et al. 2018)]{Papics2012, Buysschaert2018}. The detection of gravity modes in a strongly magnetic star makes HD~43317 unique among early-type stars, since strong magnetic fields are expected to suppress their excitation (e.g. \cite[Saio 2005; Lecoanet et al. 2017]{Saio2005, Lecoanet2017}).

\section{Forward Asteroseismic Modelling}

\begin{figure}[h]
\begin{center}
\includegraphics[width=0.9\textwidth]{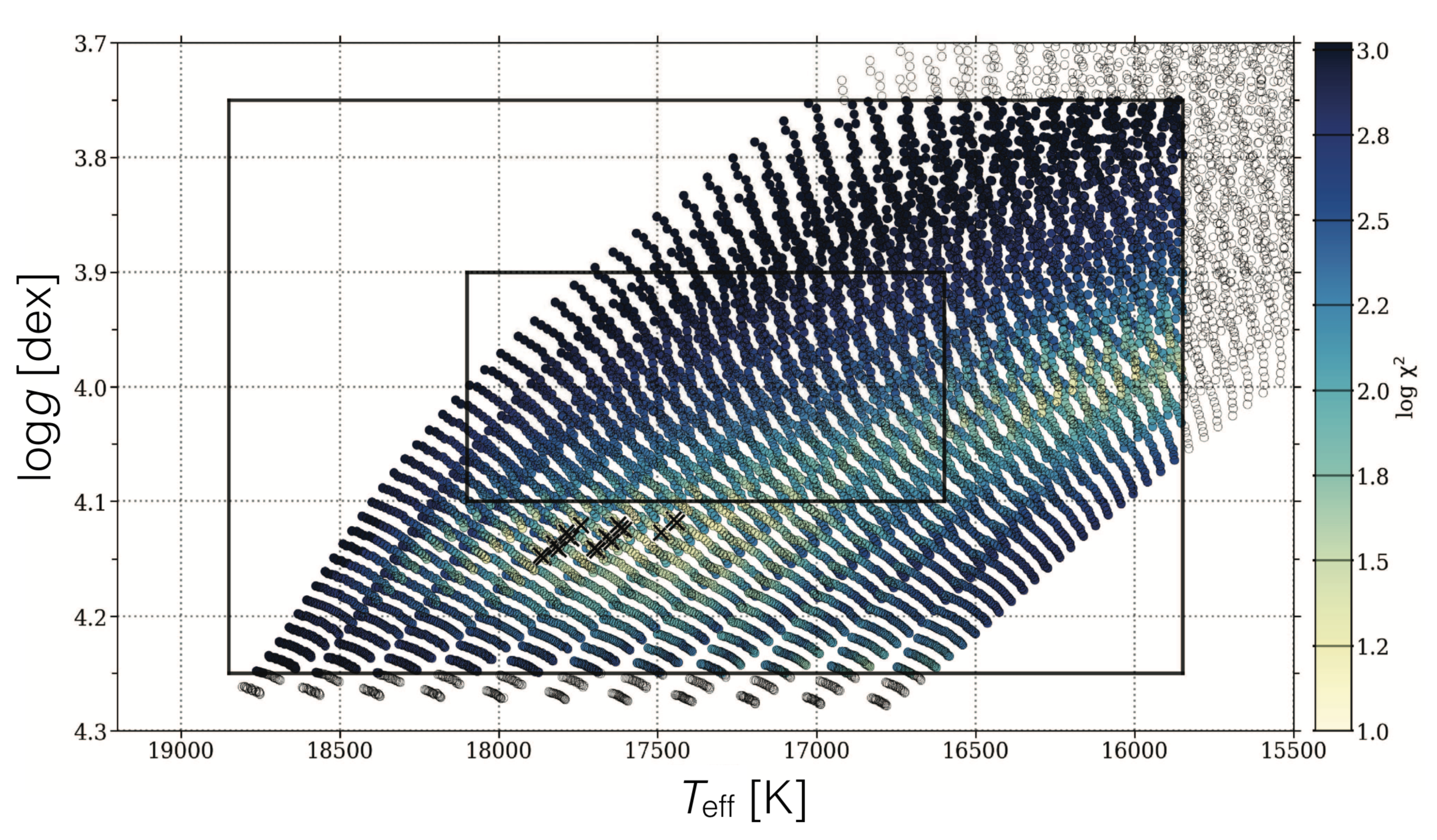}
\end{center}
\caption{Kiel diagram showing the 20 best-fitting asteroseismic structure models as black crosses for HD~43317. The 1- and 2-$\sigma$ spectroscopic confidence intervals are denoted by the two black rectangles. Each calculated stellar structure model is shown as a circle, which have been colour-coded by their resulting asteroseismic $\chi^2$ value in the forward asteroseismic modelling methodology. Figure adapted from \cite[Buysschaert et al. (2018)]{Buysschaert2018}, their figure~7.}
\label{figure: HRD}
\end{figure}

The gravity-mode pulsation mode frequencies of HD~43317 extracted from the CoRoT light curve allowed \cite[Buysschaert et al. (2018)]{Buysschaert2018} to perform forward asteroseismic modelling to ascertain the best-fitting interior structure. As a first step, the effective temperature and surface gravity from spectroscopy were used to delimit the parameter space in the Kiel diagram (see Fig.~\ref{figure: HRD}). Within this parameter space, a grid of non-rotating stellar structure models was calculated using the open-source {\sc MESA} evolution code \cite[(r8118; Paxton et al. 2011, 2013, 2015)]{Paxton2011, Paxton2013, Paxton2015}. The free parameters were the mass, $M$, the central core hydrogen content, $X_{\rm c}$, as a proxy for age on the main sequence, and the amount of convective boundary mixing (CBM) of the convective core, $f_{\rm CBM}$, assuming a diffusive exponential prescription. The metallicity was fixed at solar as indicated by the spectroscopic abundance analysis (i.e. $Z = 0.014$), and a fixed amount of mixing was set in the radiative envelope, specifically $D_{\rm env} = 10$~cm$^{2}$\,s$^{-1}$. Models were calculated from the zero-age main sequence until core hydrogen depletion (i.e. $X_{\rm c} = 0$). 

For every structure model, the adiabatic pulsation mode frequencies for dipole (i.e. $\ell=1$) and quadrupole (i.e. $\ell=2$) gravity modes and all possible azimuthal orders (i.e. $|m| \leq \ell$) were calculated using the {\sc GYRE} stellar pulsation code \cite[(Townsend \& Teitler 2013; Townsend et al. 2018)]{Townsend2013, Townsend2018}. A uniform (i.e. rigid) interior rotation profile was assumed applying the measured surface rotation period of 0.897673(4)~d and the traditional approximation for rotation (TAR) was used to include the effects of the Coriolis force on the pulsation mode frequencies. However, such an approach neglects the horizontal component and the structural deformation due to the centrifugal force, because \cite[Buysschaert et al. (2018)]{Buysschaert2018} used non-rotating structure models.

Forward asteroseismic modelling is the quantitative comparison of observed pulsation mode frequencies to those theoretically predicted from a grid of stellar structure models by use of a merit function \cite[(Aerts 2021)]{Aerts2021}. After eliminating the unlikely mode geometries, \cite[Buysschaert et al. (2018)]{Buysschaert2018} identified the best-fitting structure model and then recalculated a finer grid of {\sc MESA} models to oversample the parameter space identified by the global $\chi^2$ minimum and repeated the forward asteroseismic modelling exercise. The 20 best-fitting structure models that satisfy the $2\sigma$ confidence interval from forward asteroseismic modelling are shown in Fig.~\ref{figure: HRD}. The gravity mode frequencies of HD~43317 were identified to have radial orders between $-15 \leq n_{\rm pg} \leq -1$, and the best overall structure model had a mass of $M = 5.8^{+0.1}_{-0.2}$~M$_{\odot}$, a core hydrogen content of $X_{\rm c} = 0.54^{+0.01}_{-0.02}$ and a parameterisation of CBM to be $f_{\rm CBM} = 0.004^{+0.014}_{-0.002}$ \cite[(Buysschaert et al. 2018)]{Buysschaert2018}. 

The results of the forward asteroseismic modelling make HD~43317 particularly interesting for two reasons, both of which support the hypothesis that a strong magnetic field must exist in the deep interior. Firstly, the low amount of measured CBM is plausibly a consequence of a magnetic field suppressing convective motions from `overshooting' the Schwarzschild boundary of the convective core into the overlying radiative envelope. This is observed in magneto-hydrodynamical (MHD) simulations of magnetised core convection \cite[(e.g. Browning et al. 2004; Featherstone et al. 2009; Augustson et al. 2016)]{Browning2004, Featherstone2009, Augustson2016}. Secondly, the gravity-mode frequencies identified in HD~43317 are particularly high frequency and have atypical radial orders for gravity modes in stars with masses between $3 \lesssim M \lesssim 9$~M$_{\odot}$ (see \cite[Pedersen et al. 2021]{Pedersen2021}). This lends further support to the presence of a strong interior magnetic field, because, if present, it would interact with and suppress low-frequency gravity modes leaving only high-frequency gravity modes unaffected.

\section{Interior Magnetic Field Inference}

\begin{figure}[b]
\begin{center}
\includegraphics[width=0.49\textwidth]{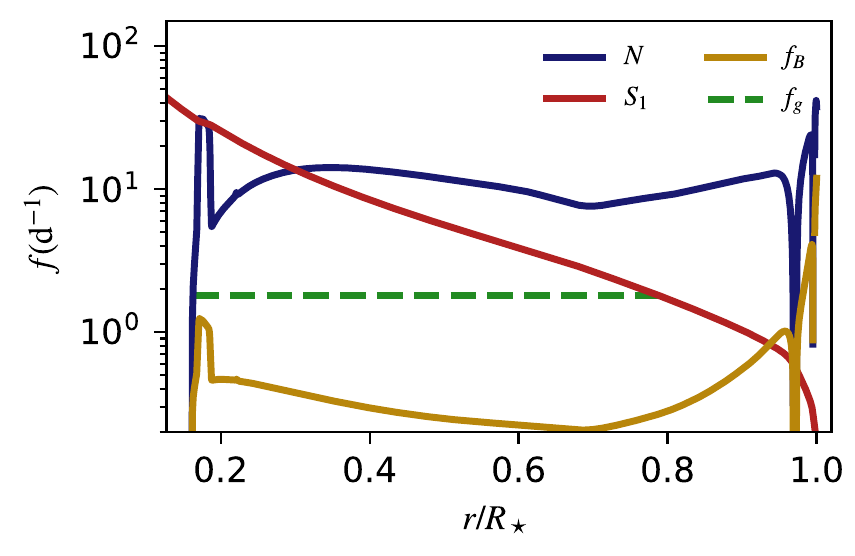}
\includegraphics[width=0.49\textwidth]{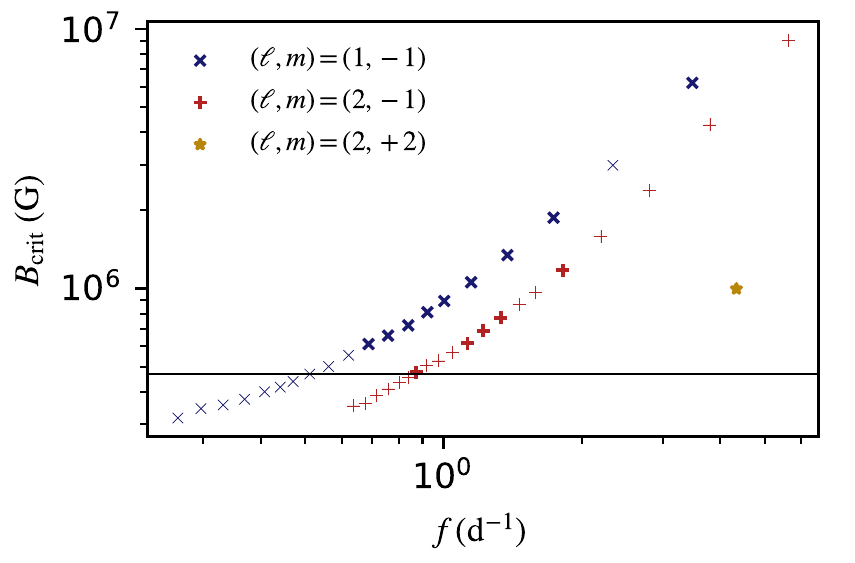}
\end{center}
\caption{Left: The best fitting structure model from forward asteroseismic modelling by \cite[Buysschaert et al. (2018)]{Buysschaert2018}, showing the Lamb ($S_{\rm l}$) and Brunt-V{\"a}is{\"a}l{\"a} ($N$) frequency profiles, the lowest observed frequency gravity mode ($f_{\rm g}$) and the magnetic interaction frequency ($f_{\rm B}$; cf. Eqn.~\ref{equation: f-B}). Right: Gravity mode frequencies as a function of critical magnetic field strength needed to suppress their propagation, in which observed gravity modes are shown as thick symbols. The horizontal line denotes the maximum permitted magnetic field strength without suppressing the lowest frequency observed gravity mode. Figures adapted from \cite[Lecoanet, Bowman \& Van Reeth et al. (2022)]{Lecoanet2022}, their figures~1 and 3.}
\label{figure: 2}
\end{figure}

To test the hypothesis of a strong magnetic field in the deep interior of HD~43317, \cite[Lecoanet, Bowman \& Van Reeth (2022)]{Lecoanet2022} performed a magneto-asteroseismic analysis to infer the critical magnetic field strength needed to suppress non-observed gravity modes. Previous work has found that gravity waves can strongly interact with magnetic fields in stellar interiors \cite[(e.g. Rogers \& MacGregor 2010; Fuller et al. 2015; Lecoanet et al. 2017)]{Rogers2010, Fuller2015, Lecoanet2017}. If a gravity wave of frequency $f$ enters a region where
\begin{equation}
f \lesssim f_{\rm B} = \frac{1}{2\pi} \sqrt{ \frac{B_{\rm r}}{\pi\rho} \frac{N\Lambda}{r} }
\label{equation: f-B}
\end{equation}
\noindent where $B_{\rm r}$ is the radial magnetic field, $\rho$ is the density, $\Lambda = \sqrt{\ell(\ell+1)}$ with $\ell$ the spherical harmonic degree of the gravity wave, and $r$ is the local radius, it converts into a magnetic (i.e. Alfv{\'e}n) wave, which in turn prevents the formation of a gravity mode (i.e. standing gravity wave). The best fitting structure model as found by \cite[Buysschaert et al. (2018)]{Buysschaert2018}, as well as the resultant magnetic interaction frequency ($f_{\rm B}$; cf. Eqn.~\ref{equation: f-B}) for HD~43317 are shown in the left panel of Fig.~\ref{figure: 2}. The lowest observed gravity mode frequency of HD~43317 is close to the value of $f_{\rm B}$ in the near-core region at $r=0.18R_{\star}$. The critical magnetic field strength required to suppress all of the non-observed low-frequency gravity modes given the structure model of HD~43317 is inferred to be $B_{\rm r} \simeq 4.7 \times 10^5$~G at $r=0.18R_{\star}$, as shown in the right panel of Fig.~\ref{figure: 2} \cite[(Lecoanet, Bowman \& Van Reeth 2022)]{Lecoanet2022}.

\begin{figure}[b]
\begin{center}
\includegraphics[width=0.49\textwidth]{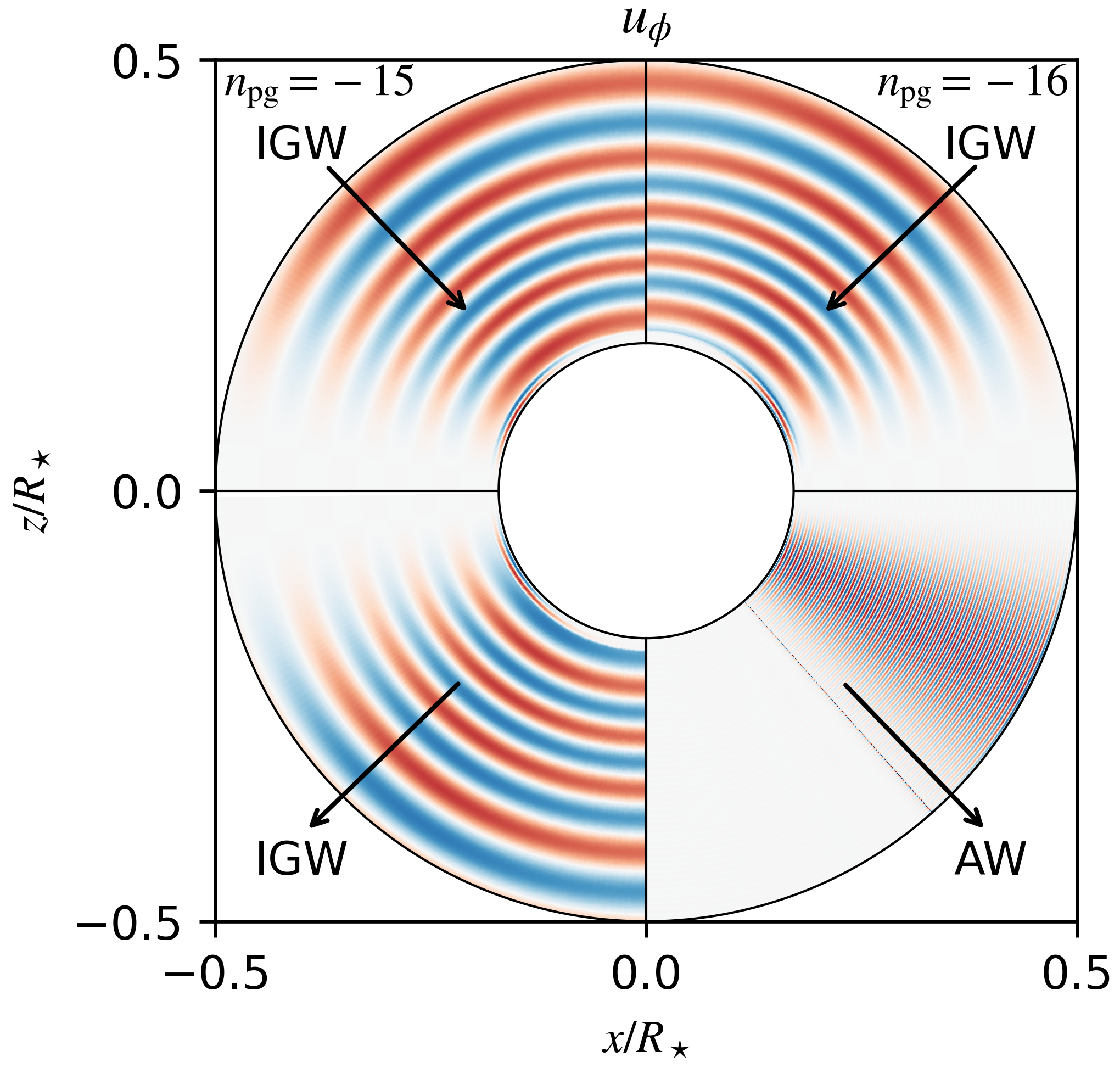}
\includegraphics[width=0.49\textwidth]{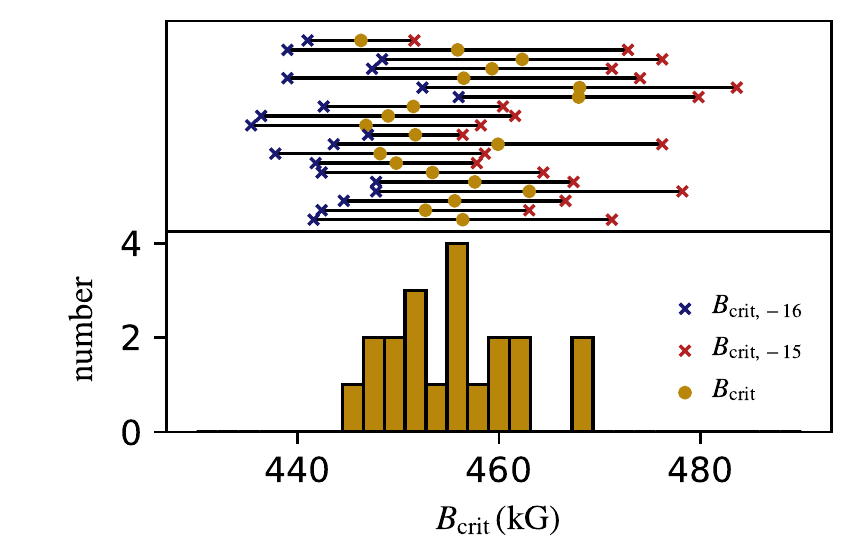}
\end{center}
\caption{Left: {\sc dedalus} calculation showing the azimuthal velocity structure for the $(\ell, m) = (2, -1)$ gravity wave with $f = 0.872$~d$^{-1}$ and $n_{\rm pg} = -15$ in the left quadrants and $f = 0.840$~d$^{-1}$ and $n_{\rm pg} = -16$ in the right quadrants. The top and bottom quadrants show the incoming and outgoing wave structure, respectively, which in the case of the right quadrants show that the gravity wave is converted into an Alfv{\'e}n wave, hence not observed. Right: Inferred magnetic field strengths in the near-core region taking the 20 best-fitting structure modelling from \cite[Buysschaert et al. (2018)]{Buysschaert2018} as input. The upper and lower limits are assuming radial orders of $n_{\rm pg} = -15$ and $-16$ for the lowest frequency observed gravity mode, respectively. Figures adapted from \cite[Lecoanet, Bowman \& Van Reeth et al. (2022)]{Lecoanet2022}, their figures~4 and 5.}
\label{figure: 3}
\end{figure}

To accurately quantify the magnetic field strength in the near-core of HD~43317, \cite[Lecoanet, Bowman \& Van Reeth (2022)]{Lecoanet2022} calculated the linear waves for a rotating and magnetised stellar model taking the best model in Fig.~\ref{figure: 2} as input and using the Wentzel-Kramers-Brillouin-Jeffreys (WKBJ) approximation with the {\sc dedalus} code \cite[(Lecoanet et al. 2019; Vasil et al. 2019; Burns et al. 2020)]{Lecoanet2019, Vasil2019, Burns2020}. Spherical geometry, a predominately dipolar magnetic field geometry and uniform interior rotation at the rate used in the prior asteroseismic analysis were assumed. A graphical representation of the results of these {\sc dedalus} calculations showing the azimuthal velocity structure of gravity modes above and below the critical magnetic field threshold is shown in the left panel of Fig.~\ref{figure: 3}. The important conclusion is that frequencies below the lowest observed $n_{\rm pg} = -15$ gravity mode strongly interact with the interior magnetic field. This places a strong constraint on the maximum magnetic field strength inside HD~43317 corroborating the observed distribution of gravity modes. To check the robustness of their results, \cite[Lecoanet, Bowman \& Van Reeth (2022)]{Lecoanet2022} propagated the uncertainties on the best-fitting structure model into the magnetic field inference and find they have only a small impact, as shown in the right panel of Fig.~\ref{figure: 3}. Thus the magnetic inference is robust to the asteroseismic uncertainties.


\section{Conclusions}

The recent results of \cite[Lecoanet, Bowman \& Van Reeth (2022)]{Lecoanet2022} place a strong constraint on the maximum magnetic field strength of $B_{\rm r} \simeq 4.7 \times 10^5$~G at $r=0.18R_{\star}$ for the magnetic pulsating early-B star HD~43317. The dominant probing region of the gravity modes is just outside the convective core (see Fig.~\ref{figure: 2}). Since HD~43317 has evolved beyond the zero-age main-sequence and its convective core has receded, the current gravity-mode cavity was previously within the convective core. The inferred near-core magnetic field strength is comparable to predictions from MHD simulations containing a fossil field and convective core dynamo in a main sequence massive star \cite[(e.g. Browning et al. 2004; Featherstone et al. 2009; Augustson et al. 2016)]{Browning2004, Featherstone2009, Augustson2016}.

HD~43317 is currently the only magnetic star pulsating in gravity modes to have undergone forward asteroseismic modelling \cite[(Buysschaert et al. 2018)]{Buysschaert2018}, but serves as a valuable proof-of-concept for magneto-asteroseismology. The ongoing TESS space mission is providing excellent time series data \cite[(Ricker et al. 2015)]{Ricker2015}, which have revealed a high pulsator fraction among massive stars (e.g. \cite[Bowman et al. 2019; Burssens et al. 2020]{Bowman2019, Burssens2020}). With future spectropolarimetric and spectroscopic campaigns, such as the CubeSpec mission \cite[(Bowman et al. 2022)]{Bowman2022}, the future of massive star asteroseismology is bright.

\section*{Acknowledgements}
\small
DMB and TVR gratefully acknowledge funding from the Research Foundation Flanders (FWO) by means of senior and junior postdoctoral fellowships with grant agreements 1286521N and 12ZB620N, respectively, and FWO long stay travel grants V411621N and V414021N, respectively. DL is supported in part by the US National Aeronautics and Space Administration (NASA) grant 80NSSC20K1280. Computations were conducted with support by the NASA High End Computing Program through the NASA Advanced Supercomputing (NAS) Division at Ames Research Center on Pleiades with allocation GIDs s2276. This research was supported in part by the National Science Foundation under grant no. NSF PHY-1748958.





\end{document}